\documentstyle[12pt]{article}

\begin{document}
\newcommand{\ti}{\theta^{1,0\,\und i}}
\newcommand{\tl}{\theta^{1,0\,\und l}}
\newcommand{\tk}{\theta^{1,0\,\und k}}
\newcommand{\tx}{\theta^{1,0}_{\und i}}
\newcommand{\ta}{\theta^{0,1\,\und a}}
\newcommand{\tb}{\theta^{0,1\,\und b}}
\newcommand{\tc}{\theta^{0,1\,\und c}}
\newcommand{\ty}{\theta^{0,1}_{\und a}}
\newcommand{\da}{D^{2,0}}
\newcommand{\db}{D^{0,2}}
\newcommand{\du}{D^{0,0}_u}
\newcommand{\dv}{D^{0,0}_v}
\newcommand{\qa}{q^{1,0\,\und a}}
\newcommand{\qb}{q^{0,1\,\und i}}
\newcommand{\qc}{q^{\und i\,\und a}}
\newcommand{\nn}{\nonumber}
\newcommand{\be}{\begin{equation}}
\newcommand{\bea}{\begin{eqnarray}}
\newcommand{\eea}{\end{eqnarray}}
\newcommand{\ee}{\end{equation}}
\newcommand{\eps}{\varepsilon}
\newcommand{\und}{\underline}
\newcommand{\p}[1]{(\ref{#1})}
\begin{titlepage}
\begin{flushright}
hep-th/0605194 \\
\end{flushright}
\vskip 1.0truecm
\begin{center}
{\Large\bf Nonlinear (4, 8, 4) Multiplet of ${\cal N}{=}8, 
d{=}1$ Supersymmetry} 
\end{center}
 \vskip 1.0truecm
\centerline{\large\bf Evgeny Ivanov}
\vskip 1.0truecm

\centerline{{\it Bogoliubov Laboratory of
Theoretical Physics, JINR,}}
\centerline{\it 141 980 Dubna, Moscow region,
Russian Federation}
\vspace{0.1cm}
\centerline{{\tt eivanov@theor.jinr.ru}}

\vskip 1.0truecm  \nopagebreak

\begin{abstract}
\noindent We construct a nonlinear version of the $d{=}1$ off-shell ${\cal N}{=}8$ multiplet 
${\bf (4, 8, 4)}\,$, proceeding from a nonlinear realization of the superconformal group  
$OSp(4^\star\vert 4)$ in the ${\cal N}{=}8\,, d{=}1$ analytic bi-harmonic 
superspace. The new multiplet is described by a double-charged analytic superfield 
$q^{1,1}$ subjected to some nonlinear harmonic constraints which are covariant under 
the $OSp(4^\star\vert 4)$ transformations. Together with the analytic 
superspace coordinates, $q^{1,1}$ parametrizes 
an analytic coset manifold of $OSp(4^\star\vert 4)\,$ 
and so is a Goldstone superfield. In any $q^{1,1}$ action the superconformal symmetry is broken, 
while ${\cal N}{=}8\,, d{=}1$ Poincar\'e supersymmetry can still be preserved. We construct the most 
general class of such supersymmetric actions and 
find the general expression for the bosonic target metric in terms of the original 
analytic Lagrangian superfield density which is thus the target geometry prepotential. 
It also completely specifies the scalar potential. The metric is conformally flat 
and, in the $SO(4)$ invariant case, is a deformation of the metric of a four-sphere $S^4\,$.              
\end{abstract}
\vfill

\newpage

\end{titlepage}

\section{Introduction}
Extended $d{=}1$ supersymmetry reveals a number of specific features which are not shared by 
its  higher-dimensional counterparts. One of such peculiarities is the existence of 
nonlinear cousins of the standard linear $d{=}1$ supermultiplets. They possess the same off-shell 
component contents, but are described by $d{=}1$ superfields subjected to some nonlinear constraints. 
As a result, the relevant $d{=}1$ supersymmetries are realized on these multiplets by  
intrinsically nonlinear transformations. While the target geometries of supersymmetric 
$d{=}1$ sigma models associated with linear multiplets were studied in detail \cite{Geom}, 
no such an exhaustive  
study was undertaken so far for the nonlinear ones. Also, it is an open question which 
realistic physical 
systems of supersymmetric quantum mechanics (SQM) can be described by $d{=}1$ models based 
on nonlinear multiplets and whether the latter can be recovered within some modified dimensional 
reduction procedure. To answer these and related questions, it is important to have more examples 
of the nonlinear multiplets and supersymmetric models constructed on their basis. 

In the ${\cal N}{=}4, d{=}1$ case nonlinear versions are known for the off-shell 
multiplets ${\bf (3,4,1)}$ \cite{IL,IKL1}, 
${\bf (2, 4, 2)}$ \cite{IKL1} and  ${\bf (4, 4, 0)}$ \cite{Pol,1,2,3}. The detailed 
structure of the corresponding 
invariant actions and interrelations between various multiplets were studied in refs.
 \cite{BBKNO}-\cite{BKS}. In the 
case of ${\cal N}{=}8, d{=}1$ supersymmetry the only nonlinear multiplet known so far 
is the appropriate counterpart of the chiral ${\cal N}{=}8$ multiplet ${\bf (2,8,6)}$ \cite{Chir8}. 

In this paper we describe one more nonlinear ${\cal N}{=}8$ multiplet, a counterpart of 
the linear multiplet 
${\bf (4, 8, 4)}$ studied in \cite{BKSu,BISu}. We make use of the manifestly 
${\cal N}{=}8$ supersymmetric language of 
the bi-harmonic ${\cal N}{=}8, d{=}1$ superspace \cite{BISu}. 
The ${\cal N}{=}4$ nonlinear multiplets 
${\bf (3,4,1)}$ and ${\bf (2,4,2)}$ were deduced in \cite{IKL1} as the Goldstone multiplets 
associated 
with nonlinear realizations of the most general ${\cal N}{=}4, d{=}1$ superconformal 
group $D(2,1; \alpha)\,$. 
The nonlinear ${\bf (4, 8, 4)}$ 
multiplet is also Goldstone multiplet, this time associated with a coset of one 
of the $d{=}1$ superconformal 
supergroups \cite{VPr}, the supergroup $OSp(4^\star\vert 4)$. 
The power of the ${\cal N}{=}8$ bi-harmonic superspace manifests 
itself  in the fact that there is no need to use the standard routine of 
nonlinear realizations, i.e. 
to start from the explicit structure relations of  $OSp(4^\star\vert 4)$, to construct 
the relevant Cartan forms, etc. 
The ${\cal N}{=}8$ multiplet in question is described by the analytic 
bi-harmonic superfield 
of the same type as 
in the case of linear ${\bf (4, 8, 4)}$ multiplet \cite{BISu}, but this superfield is now 
subjected to some 
nonlinear harmonic constraints. The precise form of these constraints is almost uniquely 
determined from the requirement 
of covariance under nonlinear $SO(5)/SO(4)$ transformations 
(belonging to $OSp(4^\star\vert 4)$) 
and nonlinear 
version of ${\cal N}{=}8, d{=}1$ Poincar\'e supersymmetry. 
The most 
characteristic novel feature of these and other transformations of the supergroup 
$OSp(4^\star\vert 4)$ on 
the analytic superspace coordinates is that they non-trivially mix these coordinates 
with the Goldstone analytic superfield. 

In Sect. 2 we recollect the pivotal features of the ${\cal N}{=}8, d{=}1$ 
bi-harmonic superspace and the superfield description of the linear ${\bf (4, 8, 4)}$ multiplet in it.  
In Sect. 3 
we define the nonlinear ${\bf (4, 8, 4)}$ multiplet within the same 
off-shell superfield framework. 
We explicitly give the analytic superspace form of the basic 
$OSp(4^\star\vert 4)$ transformations containing in their closure all other transformations. 
The simplest invariant action quadratic in 
the ${\bf (4, 8, 4)}$ superfield is constructed in Sect. 4. It already yields a non-trivial 
nonlinear sigma model in 
components. It is a deformation of the $SO(5)/SO(4)$ nonlinear sigma model. 
The action respects the invariance 
only under some subgroup of $OSp(4^\star\vert 4)$ involving 
${\cal N}{=}8, d{=}1$ Poincar\'e symmetry. It is impossible to 
set up the $OSp(4^\star\vert 4)$ invariant action in this case at all, 
because of lacking of dilaton 
among the components of the nonlinear ${\bf (4, 8, 4)}$ multiplet. 
In Sect. 4 we also construct the most general 
action of this multiplet invariant under ${\cal N}{=}8, d{=}1$ 
Poincar\'e supersymmetry and study its bosonic sector. 
We derive the nonlinear differential equation satisfied 
by the target bosonic metric and show that 
our bi-harmonic formalism supplies a natural general solution 
for this metric in terms of the analytic 
superfield Lagrangian density which thus plays the role of 
the prepotential of the relevant target 
geometry. It also completely determines the form 
of the admissible scalar potential.

\section{Bi-harmonic superspace and linear ${\cal N}{=}8, d{=}1$ multiplet (4,8,4)}

\subsection{Standard and bi-harmonic ${\cal N}{=}8$ superspaces}
The standard real ${\cal N}{=}8$\,, $d{=}1$ superspace is defined
as the following set of coordinates:
$$
{\bf R}^{(1|8)} = (\,Z\,) =
(\,t\,, \theta^{i\, \und k}\,, \theta^{a\, \und b}\,)\,.
$$
Here the indices $i$\,, $\und k$ and $a$\,, $\und b$
are doublet indices of four commuting $SU(2)$ groups forming the subgroup
$SO(4) \times SO(4)$ of the full automorphism group $SO(8)$ of
${\cal N}{=}8, d{=}1$ Poincar\'e superalgebra.
The covariant spinor derivatives are defined as
\bea
&&
D_{i\,\und k} = \frac{\partial}{\partial \theta^{i\,\und k}} +
i\,\theta_{i\, \und k}\,\partial_t\,, \quad
D_{a\, \und b} = \frac{\partial}{\partial \theta^{a\, \und b}} +
i\,\theta_{a\, \und b}\,\partial_t\,, \nn\\
&&
(D_{i\, \und k})^{\dagger} =
- \eps^{i\,l}\, \eps^{\und k\, \und n}\, D_{l\, \und n}\,, \quad
(D_{a\, \und b})^{\dagger} =
- \eps^{a\,c}\, \eps^{\und b\, \und d}\, D_{c\, \und d}\,,
\eea
and obey the following algebra:
\be
\{\,D_{i\, \und k}\,, D_{j\, \und l} \,\} = 2i\, \eps_{i\,j}\,
\eps_{\und k\, \und l}\,
\partial_t\,, \quad
\{\,D_{a\, \und b}\,, D_{c\, \und d} \,\} = 2i\, \eps_{a\,c}\,
\eps_{\und b\, \und d}\,
\partial_t\,.
\label{alg}
\ee

For the one--dimensional ${\cal N}{=}8$ supersymmetric theory
we can introduce $SU(2)\times SU(2)$ bi-harmonic superspace (HSS) with two independent
sets of harmonic variables $u^{\pm 1}_i$ and $v^{\pm 1}_a$ associated with
two different $SU(2)$ groups of the above $SO(4) \times SO(4)$ automorphism group.
As shown in \cite{BISu}, it provides the appropriate framework
for the ${\cal N}{=}8$ supersymmetric quantum mechanics associated
with the $d{=}1$ off-shell supermultiplet ${\bf (4,8,4)}$ \cite{BIKL2}.

We define the central basis of this HSS as
\be
{\bf HR}^{(1+2+2|8)} = (\,Z\,,u\,,v\,) =
{\bf R}^{(1|8)} \otimes (\,u^{\pm1}_i, v^{\pm1}_a\,)\,, \quad
u^{1i} u^{-1}_i = 1\,,\quad v^{1a} v^{-1} _a = 1\,.
\label{HSS}
\ee
The analytic basis in the same bi-harmonic SS amounts
to the following choice of coordinates:
\be
{\bf HR}^{(1+2+2|8)} = (\,X\,,u\,,v\,) = (\,t_A\,,
\theta^{\pm 1,0\, \und i}\,, \theta^{0,\pm1\, \und a}\,,
u^{\pm 1}_i\,, v^{\pm 1}_a\,)
\label{an.set}
\ee
where
$$
t_A = t + i (\ti\, \theta^{-1,0}_{\und i} + \ta\, \theta^{0,-1}_{\und a})\,, \quad
\theta^{\pm 1,0\, \und i} = \theta^{k\, \und i}\, u^{\pm 1}_k\,, \quad
\theta^{0,\pm 1\, \und a} = \theta^{b\, \und a}\, v^{\pm 1}_b\,.
$$
The analytic basis makes manifest the existence of the {\it analytic subspace} in
the bi-harmonic SS
\be
{\bf AR}^{(1+2+2|4)} = (\,\zeta\,,u\,,v\,)
= (\,t_A\,, \ti\,, \ta\,, u^{\pm 1}_i\,, v^{\pm 1}_a\,)\,,
\label{AS}
\ee
which has twice as less odd coordinates as compared to the standard ${\cal N}{=}8, d{=}1$
superspace and is closed under ${\cal N}{=}8$ supersymmetry transformations
\be
\delta t_A = 2i\left(\epsilon^{-1,0\,\und{i}}\theta^{1,0}_{\und{i}} +
\epsilon^{0,-1\,\und{a}}\theta^{0,1}_{\und{a}}\right), \quad \delta\theta^{1,0\,\und{i}} =
\epsilon^{1,0\,\und{i}}\,, \; \delta\theta^{0,1\,\und{a}} =
\epsilon^{0,1\,\und{a}}\,, \label{LinN8}
\ee
where $\epsilon^{\pm 1,0\,\und{i}} = \epsilon^{i\und{i}}u^{\pm 1}_i\,$, etc.
The existence of the analytic subspace matches the form of covariant
spinor derivatives in the analytic basis
\be
D^{1,0\, \und i} = \frac{\partial}{\partial \theta^{-1,0}_{\und i}}\,,\quad
D^{0,1\, \und a} = \frac{\partial}{\partial \theta^{0,-1}_{\und a}}
\label{sp.der}
\ee
where
\be
D^{\pm 1,0\, \und i} \equiv D^{k\, \und i}\, u^{\pm 1}_k\,,\quad
D^{0,\pm 1\, \und a} \equiv D^{b\, \und a}\, v^{\pm 1}_b\,.
\ee
The ``shortness'' of $D^{1,0 \,\und i}\,,
D^{0,1\, \und a}$ means that the analytic bi-harmonic
superfields $\Phi^{\,q,\,p}$,
\be
D^{1,0\, \und i}\, \Phi^{\,q,\,p} = D^{0,1\, \und a}\, \Phi^{\,q,\,p} = 0\,,
\label{Phi}
\ee
do not depend on
$\theta^{-1,0\, \und i}\,, \theta^{0, -1\, \und a}$ in the analytic
basis, i.e. ``live'' on the analytic subspace \p{AS}:
\be
\Phi^{\,q,\,p} = \Phi^{\,q,\,p} (\zeta, u, v)\,.
\ee

In the bi-harmonic superspace one can define two sets of mutually
commuting harmonic derivatives, each forming an $SU(2)$ algebra \cite{IvSu,BISu}.
In the analytic basis and being applied to the analytic superfields, the derivatives
with positive $U(1)$ charges, as well as the derivatives counting
the harmonic $U(1)$ charges $p,q$, read
\bea
&&
\da = \partial^{2,0} + i\, \ti \theta^{1,0}_{\und i} \partial_t\,,\quad
\du = \partial^{0,0}_u + \ti \frac{\partial}{\partial \ti}\,, \nn\\
&&
\db = \partial^{0,2} + i\, \ta \theta^{0,1}_{\und a} \partial_t\,, \quad
\dv = \partial^{0,0}_v + \ta \frac{\partial}{\partial \ta}\,,
\label{harm.der}
\eea
and
\bea
&&
\partial^{2,0} = u^{1i} \frac{\partial}{\partial u^{-1i}}\,, \quad
\partial^{0,0}_u = u^{1i} \frac{\partial}{\partial u^{1i}}
- u^{-1i} \frac{\partial}{\partial u^{-1i}}\,, \nn\\
&&
\partial^{0,2} = v^{1a} \frac{\partial}{\partial v^{-1a}}\,, \quad
\partial^{0,0}_v = v^{1a} \frac{\partial}{\partial v^{1a}} -
v^{-1a} \frac{\partial}{\partial v^{-1a}}\,.
\eea

\subsection{The linear multiplet (4,8,4)}

In the standard ${\cal N}{=}8, d{=}1$ superspace ${\bf R}^{(1|8)}$ the multiplet
with the
off-shell field content ${\bf (4,8,4)}$ is described by a real quartet
superfield $q^{\,i\, a}$ subjected to the constraints \cite{BIKL2}
\be
D^{(k\, \und k} q^{\,i)\, a} = D^{(b\, \und b} q^{\,k\, a)} = 0\,,
\label{q11constr}
\ee
where symmetrization is understood for the doublet indices of the same $SU(2)$ group.

In the superspace ${\bf HR}^{(1+2+2|8)}$  the same multiplet ${\bf (4,8,4)}$ can be
described by a real analytic ${\cal N}{=}8$ superfield
$q^{1,1}(\zeta, u, v)$ subjected to the harmonic constraints
\be
\da q^{1,1} = 0\,,\quad \db q^{1,1} = 0\,,
\label{hc}
\ee
which in the central basis imply
\be
q^{1,1} = q^{i\,a}u^1_i v^1_a\,. \nn\\
\ee
Then $q^{ia}$ satisfies the constraints \p{q11constr} as a consequence
of the constraints \p{Phi}
\be
D^{1,0\, \und i}\, q^{1,1} = D^{0,1\, \und a}\, q^{1,1} = 0\,.
\label{qanalG}
\ee

The analytic basis solution of the harmonic constraints \p{hc} is given by
\bea
q^{1,1} \!&=&\! f^{i\,a}u^1_i v^1_a + \ti \psi^a_{\und i} v^1_a
+ \ta \psi^i_{\und a} u^1_i
- i\, (\theta^{1,0})^2 \partial_t f^{i\,a}u^{-1}_i v^1_a
- i\, (\theta^{0,1})^2 \partial_t f^{i\,a}u^1_i v^{-1}_a \nn\\
\!&+&\!
\ti \ta F_{\und i\, \und a}
- i\, \ti (\theta^{0,1})^2 \partial_t \psi^a_{\und i} v^{-1}_a
- i\, \ta (\theta^{1,0})^2 \partial_t \psi^i_{\und a} u^{-1}_i \nn\\
\!&-&\!
(\theta^{1,0})^2 (\theta^{0,1})^2 \, \partial_t^2 f^{i\,a}u^{-1}_i v^{-1}_a
\label{sol}
\eea
where $(\theta^{1,0})^2 = \theta^{1,0\, \und k}\,\theta^{1,0}_{\und k}$\,,
$(\theta^{0,1})^2 = \theta^{0,1\, \und a}\,\theta^{0,1}_{\und a}\,$.
The independent component $d{=}1$ fields
$f^{ia}, \psi^a_{\und i}, \psi^i_{\und a}$ and $F_{\und i\, \und a}$
form the ${\cal N}{=}8$ off-shell multiplet ${\bf(4, 8, 4)}$.

The general {\it off-shell} action of $n$ such superfields $q^{1,1\, M}$
$(M = 1,2,...n)$ can be written as 
\be
S^{gen} = \int \mu^{(-2,-2)}\, {\cal L}^{2,2} (q^{1,1\, M},u,v)
\label{s1gen}
\ee
where
\be
\mu^{(-2,-2)} = dt du\, dv\,d^2\theta^{1,0}\, d^2\theta^{0,1}
\label{measure1}
\ee
is the analytic superspace integration measure normalized as
$$
\int d^2\theta^{1,0}\, d^2\theta^{0,1} (\theta^{1,0})^2 (\theta^{0,1})^2 = 1\,.
$$
The analytic superfield Lagrangian ${\cal L}^{2,2}$ can bear an arbitrary
dependence on its arguments, the only restriction being a compatibility
with its external charges $(2,2)$\,.
The free action is 
\be
S^{free} = \int \mu^{(-2,-2)}\, q^{1,1\, M} q^{1,1\, M}\,.
\label{free}
\ee
Using (\ref{sol})\,, one finds the component form of the action (\ref{free})
(for one $q^{1,1}$)
\be
S^{free} = \frac{1}{2}\,\int dt\,
\Big \{\, \partial_t f^{i\,a}\,\partial_t f_{i\,a}
+ \frac{i}{2}\,\Big (\,\psi^{\und i\,a}\, \partial_t \psi_{\und i\,a} +
\psi^{i\, \und a}\, \partial_t \psi_{i\, \und a}\, \Big )
+ \frac{1}{4}\, F^{\und i}_{\und a}\, F_{\und i}^{\und a}\, \Big \}\,.
\ee
The bosonic part of the general action (\ref{s1gen}) for one $q^{1,1}$ reads
\bea
S^{n=1}_{bos} = \frac{1}{2} \,\int dt\, \Big \{
G(f)\,\partial_t f^{i\,a}\,\partial_t f_{i\,a}
+ \frac{1}{4}\,G(f)\, F^{\und i}_{\und a}\, F_{\und i}^{\und a}\Big \}
\label{Scomp1}
\eea
where
\bea
&& G(f) = \int du\,dv\, g(f^{1,1}\,,u\,,v)\,, \quad
g(f^{1,1}\,,u\,,v) = \left. \frac{\partial^2 {\cal L}^{2,2}}{\partial q^{1,1}\,
\partial q^{1,1}}
\right|_{\theta = 0}\,, \label{g} \\
&& \left. q^{1,1} \right|_{\theta = 0} = f^{1,1} = f^{i\, a}(t) u^1_i v^1_a\,.\nonumber
\eea
After eliminating the auxiliary field by its equation of motion
\be
F^{\und k}_{\und b} = 0
\label{f}
\ee
(modulo some fermionic terms)
one obtains the on--shell form of the action (\ref{Scomp1})
\bea
S^{gen} \!&=&\! \frac{1}{2} \, \int dt\,
G(f)\,\partial_t f^{i\,a}\,\partial_t f_{i\,a}\,.
\label{onshell1}
\eea
The function $G(f)$ defined in (\ref{g}) satisfies the four--dimensional Laplace equation
\be
\bigtriangleup G(f) = 0\,, \quad
\bigtriangleup = \frac{\partial^2}{\partial f^{i\,a}\, \partial f_{i\,a}}\,
\label{Lap}
\ee
which follows from the definition of $G(q)$\,. 

For any number of the ${\bf (4,8,4)}$ multiplets one deals with the same 
bosonic target HKT (hyper-K\"ahler with torsion) geometry as in the case
of twisted ${\cal N}{=}(4,4), d{=}2$ multiplets \cite{HKT}.

\setcounter{equation}{0}

\section{Nonlinear (4,8,4) supermultiplet} 

In ${\cal N}{=}(4,4)$, $d{=}2$ bi-harmonic superspace the requirement of preserving the flat
form of the harmonic derivatives $D^{2,0}$, $D^{0,2}$ uniquely selects
the infinite-dimensional ``large'' ${\cal N}{=}4$ superconformal groups
(both in the left and right
light cone sectors) as the most general coordinate groups meeting this requirement.
The same requirement in the $d{=}1$ version of bi-harmonic superspace, together
with the demand of covariance of the defining $q^{1,1}$ constraints \p{hc},
pick up a supergroup which does
not coincide with any known $d{=}1$ superconformal group \cite{VPr}.
It is some ${\cal N}{=}8$ superextension of the Heisenberg group ${\bf h}(2)$ with
an operator central charge \cite{BISu}.

On the other hand, one can expect that the multiplet $({\bf 4, 8, 4})$,
along the lines of ref.~\cite{IKL1,BIKL2},
could be treated as the Goldstone multiplet parametrizing the appropriate coset
of the ${\cal N}{=}8, d{=}1$ superconformal group $OSp(4^*|4)$ \cite{BIKL2}.
This supergroup admits a realization on the coordinates $Z$ of the
standard ${\cal N}{=}8, d{=}1$ superspace ${\bf R}^{(1|8)}$ and on the constrained superfield
$q^{ia}(Z)$ representing the multiplet $({\bf 4,8,4})$ in ${\bf R}^{(1|8)}$
(see \p{q11constr}). So one can wonder why this superconformal group
does not show up
in the analytic superspace description of the multiplet $({\bf 4, 8, 4})$, i.e. why
it is absent in the set of coordinate transformations preserving the flat form of
$D^{2,0}, D^{0,2}$. As explained in \cite{BISu}, the reason is that passing to
the bi-harmonic extension of ${\bf R}^{(1|8)}$ reduces the general R-symmetry group
$SO(8)$ of ${\bf R}^{(1|8)}$  down to its subgroup $SO(4)\times SO(4)$, while no
${\cal N}{=}8, d{=}1$ superconformal
groups with such R-symmetry exist \cite{VPr}. In particular, R-symmetry subgroup of
$OSp(4^*|4)$ is $USp(4)\times SU(2) \sim SO(5)\times SU(2)$. Hence,
in the $d{=}1$ bi-harmonic superspace it is impossible to realize any standard
${\cal N}{=}8$ superconformal group, under the assumption that the corresponding
R-symmetry group acts {\it linearly} on the harmonic variables
$u^{\pm 1}_i, v^{\pm 1}_a$. It was also noticed in \cite{BISu} that, given Goldstone
$({\bf 4,8, 4})$ multiplet, with physical bosons parametrizing the R-symmetry
coset $SO(5)/SO(4)$, the R-symmetry $SO(5)$ should act on the harmonic variables
by the transformations which are {\it nonlinear} in these physical bosonic
fields. This extends to the whole $OSp(4^*|4)$ group which was conjectured to admit
a realization in the analytic bi-harmonic superspace, such that the corresponding
coordinate variations involve the superfield $q^{1,1}$ itself. Here we show
that this hypothesis is true. We explicitly present the $SO(5)/SO(4)$
transformations and transformations of the ${\cal N}{=}8$ Poincar\'e supersymmetry.
Surprisingly, the latter are also nonlinear in $q^{1,1}$.

Taking into account that $OSp(4^\star |4)$ involves only three commuting $SU(2)$ groups,
the harmonic superspace description of the nonlinear ${\bf (4,8,4)}$ multiplet, as opposed
to the description of the linear one,  should respect only
three out of four $SU(2)$ symmetries realized in the analytic superspace.
To have a correspondence with the linear multiplet, it is natural to assume that two $SU(2)$
acting on the harmonic variables still remain unbroken. So we are led to identify two
$SU(2)$ groups acting on the external doublet indices $\und{i}$ and $\und{a}$ of the
analytic Grassmann coordinates. We also should always keep the balance of the harmonic
$U(1)$ charges. The nonlinear transformations of the ${\cal N}{=}8$
supersymmetry still preserving the Grassmann bi-harmonic analyticity are then as follows:
\bea
&& \delta t = 2i\left(\epsilon^{-1,0\,\und{i}} \theta^{1,0}_{\und{i}} +
\epsilon^{0, -1\,\und{i}} \theta^{0,1}_{\und{i}}\right), \;
\delta \theta^{1,0\,\und{i}} =
\epsilon^{1,0\,\und{i}} + q^{1,1}\epsilon^{0,-1\,\und{i}}\,,
\; \delta\theta^{0,1\,\und{i}} =  \epsilon^{0,1\,\und{i}} +
q^{1,1}\epsilon^{-1,0\,\und{i}}\,, \nn \\
&& \delta q^{1,1} \simeq q^{1,1}{}'(\zeta',u)
- q^{1,1}(\zeta, u) = 0\,,\label{NonlN8}
\eea
which should be compared with \p{LinN8}. It is easy to see that these
modified ${\cal N}{=}8$ supersymmetry transformations
close on the $t$-translations in the same way as the standard linear
${\cal N}{=}8$ ones. They manifestly preserve the analytic superspace. The analytic 
measure \p{measure1} transforms as
\be
\delta \mu^{(-2,-2)} = - \left(\partial_{1,0\,\und{i}}q^{1,1}\epsilon^{0,-1\,\und{i}}
+ \partial_{0,1\,\und{i}}q^{1,1}\epsilon^{-1,0\,\und{i}} \right)
\mu^{(-2,-2)}\,. \label{Tranmu}
\ee
From this transformation law it follows, in particular, that the analytic superspace
integral of any power of $q^{1,1}$ is invariant up to a total derivative in the integrand.

The flat harmonic derivatives \p{harm.der} are obviously not covariant with respect to the
transformations \p{NonlN8}. It is rather straightforward to find their
fully covariant analogs
\bea
\hat{D}^{2,0} = D^{2,0} + q^{1,1}\left( \theta^{1,0\,\und{i}}\partial_{0,1\,\und{i}} +
q^{1,1} \partial^{0,-2}\right), \quad
\hat{D}^{0,2} = D^{0,2} + q^{1,1}\left( \theta^{0,1\,\und{i}}\partial_{1,0\,\und{i}} +
q^{1,1} \partial^{-2,0}\right), \label{HatD}
\eea
where $\partial^{-2,0} = u^{-1}_i\frac{\partial}{\partial u^1_i}\,$ and
$\partial^{0, -2} = v^{-1}_a\frac{\partial}{\partial v^1_a}\,$. One can check that
these operators are invariant under \p{NonlN8} up to terms
containing $\hat{D}^{2,0}q^{1,1}$ and $\hat{D}^{0,2}q^{1,1}$. Then, as the natural
generalization of the linear case constraints \p{hc}, we are led to impose on $q^{1,1}$
the following constraints:
\be
\hat{D}^{2,0} q^{1,1} = \hat{D}^{0,2} q^{1,1} = 0\,. \label{HatConstr}
\ee
With taking them into account, one easily finds that $\delta \hat{D}^{2,0} = 
\delta \hat{D}^{2,0} = 0$ under \p{NonlN8}. One should still check the integrability condition
\be
[\hat{D}^{2,0},\hat{D}^{0,2}] q^{1,1} = 0\,. \label{IntNonl}
\ee
Commuting the derivatives \p{HatD} with each other and making use of \p{HatConstr}, we find
\be
[\hat{D}^{2,0},\hat{D}^{0,2}] = (q^{1,1})^2 \left(D^0_u - D^0_v\right),
\ee
whence \p{IntNonl} directly stems.

The constraints \p{HatConstr} defining the nonlinear version of the ${\bf (4,8,4)}$
multiplet are off-shell like their linear counterparts \p{hc} and go over into the latter,
if we rescale $q^{1,1}$ by some contraction parameter and then send this parameter
to zero. In this limit \p{HatD} and \p{NonlN8} turn into their standard linear
counterparts. In what follows we keep this parameter equal to 1, having in mind
that it can be re-introduced at any step by the proper rescaling.

Before solving the constraints \p{HatConstr} and constructing invariant
actions, let us show that \p{HatConstr} are covariant under a special realization
of $SO(5)$ transformations. We consider only the transformations belonging
to the coset $SO(5)/SO(4)$, with the constant parameters $b^{ia}\,$.
Defining $b^{\pm 1, \pm 1} = b^{ia}u^\pm_iv^\pm_a\,$,
$b^{\pm 1, \mp 1} = b^{ia}u^\pm_iv^\mp_a\,$, one can check that the requirement of
preserving both the Grassmann harmonic analyticity and the constraints
\p{HatConstr} fixes the corresponding trnsformations, up to a rescaling of $b^{ia}$,
as
\bea
&& \delta u^1_i = 2 (b^{1,-1}q^{1,1})u^{-1}_i\,, \;
\delta v^1_a = 2 (b^{-1,1}q^{1,1})v^{-1}_a\,, \;
\delta u^{-1}_i = \delta v^{-1}_a = 0\,, \label{HarmCoset} \\
&& \delta \theta^{1,0\,\und{i}} = (b^{-1,-1}q^{1,1}) \theta^{1,0\,\und{i}} +
b^{1,-1} \theta^{0,1 \,\und{i}}\,,\;\;
\delta \theta^{0,1\,\und{i}} = (b^{-1,-1}q^{1,1}) \theta^{0,1\,\und{i}} +
b^{-1,1} \theta^{1,0\,\und{i}}\,, \label{ThetaCoset} \\
&& \delta t = 2i b^{-1,-1}\,
\theta^{0,1\,\und{i}} \theta^{1,0}_{\und{i}}\,, \label{tCoset} \\
&& \delta q^{1,1} = b^{1,1} + b^{-1,-1}(q^{1,1})^2\,, \label{CosetTran} \\
&& \delta \hat{D}^{2,0} = -2 (b^{1,-1}q^{1,1})D^0_u\,, \quad
\delta \hat{D}^{0,2} = -2 (b^{-1,1}q^{1,1})D^0_v\,. \label{HatCoset}
\eea
Note that, while checking \p{HatCoset}, one should take into account
the constraints \p{HatConstr}, which, on their own, can be easily checked to
be covariant under \p{HarmCoset} - \p{HatCoset}.  The non-symmetric
transformations of the harmonic variables as in \p{HarmCoset} and those
of the covariant harmonic derivatives as in \p{HatCoset} are
typical for the realizations of superconformal group in the analytic harmonic
superspaces \cite{HSS,IvSu}. Indeed, the $SO(5)/SO(4)$ transformations 
\p{HarmCoset} - \p{HatCoset} are a part of the $d{=}1$ superconformal group 
$OSp(4^\star|4)\,$. The transformation
\p{CosetTran} contains a shift by $b^{1,1} = b^{ia}u^1_iv^1_a\,$, which means that
four bosonic physical fields in $q^{1,1}$ are Goldstone fields belonging to 
the coset $SO(5)/SO(4)\,$. By computing the Lie brackets of the transformations
\p{HarmCoset} - \p{HatCoset},  we can find the explicit realization of the stability
subgroup $SO(4)\,$ on the analytic coordinates and the Goldstone superfield
$q^{1,1}$
\bea
&& \delta u^1_j = \left[\lambda^{ik}u^1_iu^1_k +
(\lambda^{ab}v^{-1}_av^{-1}_b)(q^{1,1})^2 \right] u^{-1}_j\,, \;\;
\delta u^{-1}_j = 0\,, \nn \\
&& \delta v^1_c = \left[\lambda^{ab}v^1_av^1_b +
(\lambda^{ik}u^{-1}_iu^{-1}_k)(q^{1,1})^2 \right] v^{-1}_c\,,\; \; \delta v^{-1}_c = 0\,,
\label{SUtranH} \\
&& \delta \theta^{1,0\,\und{i}} = \lambda^{(ik)}u^{-1}_iu^1_k\,\theta^{1,0\,\und{i}} +
\lambda^{(ab)}v^{-1}_av^{-1}_b q^{1,1}\theta^{0,1\,\und{i}}\,, \nn \\
&& \delta \theta^{0,1\,\und{i}} = \lambda^{(ab)}v^{-1}_av^1_b\,\theta^{0,1\,\und{i}} +
\lambda^{(ik)}u^{-1}_iu^{-1}_k q^{1,1}\theta^{1,0\,\und{i}}\,, \nn \\
&& \delta t = i\lambda^{(ik)}u^{-1}_iu^{-1}_k (\theta^{1,0})^2 +
i \lambda^{(ab)}v^{-1}_av^{-1}_b (\theta^{0,1})^2\,, \label{SUcoord} \\
&& \delta q^{1,1} =
\left(\lambda^{ik}u^1_iu^{-1}_k  + \lambda^{ab}v^1_av^{-1}_b\right) q^{1,1}\,.
\label{SUtranq}
\eea
Here $\lambda^{(ik)}$ and $\lambda^{(ab)}$ are the parameter of two mutually
commuting $SU(2)$ constituents of $SO(4)\,$. Note that the Lie brackets of the
nonlinear ${\cal N}{=}8$ supersymmetry with these $SO(4)$ transformations
are again of the form \p{NonlN8}, with the proper bracket spinor parameters. The
$SO(5)/SO(4)$ transformations \p{HarmCoset} - \p{CosetTran} mix the $\epsilon^{i\und{i}}$
and $\epsilon^{a\und{i}}$ supersymmetries.

It is easy to show the covariance of \p{HatConstr} under the full superconformal group
$OSp(4^\star|4)\,$. We explicitly present only $d{=}1$ conformal
boosts with the parameter $a$
\bea
&& \delta t = a t^2\,, \quad \delta \theta^{1,0\,\und{i}} = a t\, \theta^{1,0\,\und{i}}\,,
\;\;\delta \theta^{0,1\,\und{i}} = a t\, \theta^{0,1\,\und{i}}\,, \nn \\
&& \delta u^1_i = -2i a(\theta^{1,0})^2 u^{-1}_i\,, \quad
\delta v^1_c = -2i a(\theta^{0,1})^2 v^{-1}_c\,, \quad \delta u^{-1}_i =
\delta v^{-1}_c = 0\,, \nn \\
&& \delta \hat{D}^{2,0} = 2i a(\theta^{1,0})^2 D^0_u\,,\quad
\delta \hat{D}^{0,2} = 2i a(\theta^{0,1})^2 D^0_v\,, \label{conf} \\
&& \delta q^{1,1} = -2i a (\theta^{1,0}\theta^{0,1})\,. \label{confq}
\eea
It is sufficient to check covariance only under these transformations, since the conformal
supersymmetry appears in the commutator of the conformal boosts with
the ${\cal N}{=}8, d{=}1$ Poincar\'e supersymmetry \p{NonlN8}. Then the covariance
under conformal
supersymmetry and, hence, the full superconformal group $OSp(4^\star|4)$, follows
from the covariance under \p{NonlN8} and \p{conf}, \p{confq}. In particular, the
$SO(5)$ transformations \p{HarmCoset} - \p{CosetTran} and \p{SUtranH} - \p{SUtranq},
as well as one more $SU(2)$ symmetry realized on the
doublet indices of the analytic Grassmann coordinates $\theta^{1,0\,\und{i}},
\theta^{0,1\,\und{i}}$, are contained in the closure of the conformal and ordinary
${\cal N}{=}8$ supersymmetries. It is instructive to present the transformation law of
$q^{1,1}$ under conformal ${\cal N}{=}8$ supersymmetry
\be
\delta q^{1,1} = 2i\left[ (\eta^{1,0}\theta^{0,1}) + (\theta^{1,0}\eta^{0,1}) +
q^{1,1} (\eta^{0,-1}\theta^{0,1}) + q^{1,1}(\theta^{1,0}\eta^{-1,0})\right], \label{cSUSY}
\ee
where
$\eta^{\pm 1, 0\,\und{i}} = \eta^{i \und{i}}u^{\pm 1}_i\,$,
$\eta^{0, \pm 1\,\und{i}} = \eta^{a \und{i}}v^{\pm 1}_a$ and $ \eta^{i \und{i}},
\eta^{a \und{i}}$ are the corresponding Grassmann parameters.

The presence of the shift parts in the transformations \p{confq} and \p{cSUSY}
signals that the generators of both the conformal boosts and
conformal ${\cal N}{=}8$ supersymmetry
belong to the coset part of the nonlinear realization of $OSp(4^\star|4)$ we
are dealing with, like the  generators of the $SO(5)/SO(4)$ transformations.
Correspondingly, the $SU(2)$ singlet
part of the auxiliary field in $q^{1,1}$ and all eight physical fermions are Goldstone
fields associated with these generators, in a close analogy to four physical bosons
which are Goldstone fields associated with the coset $SO(5)/SO(4)$ generators.
The set of transformations which are homogeneously realized on fields 
includes Poincar\'e ${\cal N} {=}8$
supersymmetry, $SO(4) \subset SO(5)$, $SU(2)$ which acts on the underlined doublet
indices and dilatations. The latter also appear in the closure of Poincar\'e and conformal
supersymmetries and act as the proper rescalings of $t$ and analytic Grassmann
coordinates. The superfield $q^{1,1}$ (and, respectively, the physical bosons
with which $q^{1,1}$ starts) has the zero dilatation weight. The absence of the dilaton 
(i.e. a field with a non-zero dilatation weight) among the physical bosonic fields 
is the indication that it is impossible to construct conformally invariant
(and $OSp(4^\star|4)$ invariant) actions for the nonlinear ${\bf (4,8,4)}$
multiplet, in contradistinction to its linear variant for which scale invariant
actions exist \cite{BISu}. This ``no-go'' theorem also follows from the fact that
the analytic superspace integration measure possesses a non-zero dilatation weight
(its dimension is 1) and there is no way to compensate its scale transformation 
since $q^{1,1}$ has the weight 0. So in order to construct superconformally invariant
actions of the nonlinear $q^{1,1}$, one needs at least one extra ${\cal N}{=}8$
multiplet with the dilaton among its components.
However, some symmetries from $OSp(4^\star|4)$ can be still
preserved and, first of all, the ${\cal N}{=}8$ Poincar\'e supersymmetry \p{NonlN8}.

In the next Section we will construct the most general ${\cal N}{=}8$ supersymmetric action
of the superfield $q^{1,1}$ defined by the nonlinear constraints \p{HatConstr} and
study the structure of its bosonic sector. Besides the sigma-model type action
we will also construct the general potential-type invariant
which yields a non-trivial scalar potential upon eliminating the auxiliary fields.

We close this Section with two comments.

\begin{itemize}
\item The nonlinear realization of $OSp(4^\star|4)$ constructed above
non-trivially mixes the analytic superspace coordinates and the superfield $q^{1,1}\,$.
The latter can be treated as one more analytic coordinate extending the original
analytic bi-harmonic superspace to an invariant coset space of the supergroup
$OSp(4^\star|4)$. In this respect the considered system supplies a nice example of
the ``democracy'' between coordinates and fields. In the extended analytic superspace
$(t, \theta^{1,0\,\und{i}},\theta^{0,1\,\und{i}}, q^{1,1}, u^{\pm 1}_k, v^{\pm 1}_c)$
 the original (purely coordinate) analytic superspace $(\zeta, u, v)$ specifies a
$(1+2+2|4)$ dimensional hypersurface. The Goldstone superfield  $q^{1,1}(\zeta, u, v)$
subjected to the covariant conditions \p{HatConstr} describes the embedding of
this hypersurface into the extended analytic superspace.

\item The covariantized harmonic derivatives \p{HatD} look as a particular case of the
harmonic derivatives which are pertinent to ${\cal N}{=}8, d{=}1$
supergravity \cite{BISu}. Their form \p{HatD} corresponds
to the particular choice of analytic vielbeins in the ${\cal N}{=}8$ supergravity harmonic
derivatives as functions of the constrained superfield $q^{1,1}\,$. This is an indication
that the considered system could be reproduced as a special limit of
the  ${\cal N}{=}8, d{=}1$ supergravity. 
\end{itemize}

\setcounter{equation}{0}
\section{Invariant actions}

As already mentioned, the general analytic superspace action
\be
S^{nonl}_{gen} = \int \mu^{(-2,-2)}\, {\cal L}^{2,2}(q^{1,1}, u, v) \label{NONL}
\ee
is invariant, up to a total derivative, with respect to the ${\cal N}{=}8$ supersymmetry
transformations \p{NonlN8}. This follows from the property that $q^{1,1}(\zeta, u, v)$
transforms as a scalar under \p{NonlN8} and the transformation rule of the integration
measure \p{Tranmu}. It turns out that, due to the nonlinearity of the $q^{1,1}$ constraints
\p{HatConstr}, even the ``free'' action
\be
S_{2}^{nonl} =  \int \mu^{(-2,-2)}\, q^{1,1}q^{1,1}\label{NONL2}
\ee
yields a non-trivial sigma-model type action, in contradistinction to the case
of the linear ${\bf (4,8,4)}$ multiplet. The superfield action \p{NONL2} can be checked to
respect the homogeneously realized $SO(4)$ symmetry 
\p{SUtranH} - \p{SUtranq}, while
in the general action \p{NONL} this symmetry is broken. The action \p{NONL}
and its particular case \p{NONL2} respect the third $SU(2)$ symmetry acting 
on the underlined indices of the Grassmann coordinates and component fields.

One can construct, as in the case of the linear ${\bf (4,8,4)}$ multiplet \cite{BISu},
one more invariant 
\be
S^{nonl}_{pot} = \int  \mu^{(-2,-2)}\, \theta^{1,0\,\und{i}}\theta^{0,1\,\und{k}}
C_{\und{i}\und{k}} q^{1,1}\,. \label{Pot}
\ee
Despite the explicit presence of Grassmann coordinates, the action \p{Pot} can be
checked to be invariant under the nonlinear transformations \p{NonlN8}, 
provided that
\be
C_{\und{i}\und{k}} = 2im\, \varepsilon_{\und{i}\und{k}}\,, \label{Cond5}
\ee
where $m$ is a constant of dimension 1 (the specific normalization
has been chosen for further convenience). The proof of invariance is based on representing
$\epsilon^{1,0\,\und{i}} = D^{2,0}\epsilon^{-1,0\,\und{i}}\,$,
$\epsilon^{0,1\,\und{i}} = D^{0,2}\epsilon^{0,-1\,\und{i}}\,$, integrating by parts
with respect to $D^{2,0}$, $D^{0,2}$ and making use of the constraints \p{HatConstr}.
In a similar way, the action \p{Pot} with the condition \p{Cond5} can be proved
to be invariant also
under the $SO(4)$ transformations \p{SUtranH} - \p{SUtranq}. The invariance under
the third $SU(2)$ acting on the underlined indices is obvious. In components, the
invariant \p{Pot} yields a term linear in an auxiliary field and, as in the case
of the linear ${\bf (4,8,4)}$ multiplet, produces scalar potentials after elimination of
this field by its equation of motion.

In order to pass to the components in the actions\p{NONL}, \p{NONL2}, \p{Pot} and to 
reveal the relevant bosonic target metric, we firstly need to solve the constraints \p{HatConstr}.

\subsection{Solving the constraints}
In what follows we will limit our consideration to the bosonic sector, with
all fermionic fields omitted. So we start with the following component expansion of
$q^{1,1}$
\be
q^{1,1}(\zeta, u, v) = f^{1,1} + (\theta^{1,0})^2 g^{-1,1} + (\theta^{0,1})^2 g^{1,-1} +
\theta^{1,0\,\und{i}}\theta^{0,1\,\und{k}} F_{\und{i} \und{k}} +
(\theta^{1,0})^2(\theta^{0,1})^2 d^{-1,-1}\,, \label{Theta}
\ee
where all component fields are functions of $t$ and harmonics
$u^{\pm 1}_i, v^{\pm 1}_c \,$. The superfield constraints \p{HatConstr} amount
to a set of harmonic differential equations for these
component fields. These equations fix the harmonic dependence of the fields and
express them in terms of the ${\bf (4, 4)}$ bosonic fields $f^{ia}(t)$ and
$F_{\und{i} \und{k}}(t)$ which form the bosonic subset of the off-shell nonlinear
${\bf (4,8,4)}$ multiplet. These basic 8 bosonic fields (and 8 fermionic
fields in the general case) appear as integration constants in the solutions
of the harmonic equations.

The component harmonic equations implied by the constraints \p{HatConstr}
are as follows:
\bea
&& \mbox{(a)}\,\nabla^{2,0}f^{1,1}= 0\,, \quad
\mbox{(b)}\, \nabla^{0,2}f^{1,1}= 0\,,\label{A1}\\
&&\mbox{(a)}\,{\cal D}^{2,0}g^{-1,1} +
i\partial_t{f}^{1,1} - \frac{1}{2} f^{1,1} F = 0\,, \quad
\mbox{(b)}\,{\cal D}^{0,2}g^{-1,1} = 0\,,
\label{B}\\
&&\mbox{(a)}\,{\cal D}^{2,0}g^{1,-1} = 0\,,\quad
 \mbox{(b)}\,{\cal D}^{0,2}g^{1,-1} +
i\partial_t{f}^{1,1} - \frac{1}{2} f^{1,1} F = 0\,, \label{C}\\
&& \mbox{(a)} \,{\cal D}^{2,0}F_{\und{i}\und{k}}
+ \varepsilon_{\und{i}\und{k}} f^{1,1}g^{1,-1} = 0\,, \quad
\mbox{(b)} \, {\cal D}^{0,2}F_{\und{i}\und{k}}
+ \varepsilon_{\und{i}\und{k}} f^{1,1}g^{-1,1} = 0\,, \label{D}\\
&& \mbox{(a)} {\cal D}^{2,0}d^{-1,-1}
+ i\partial_t{g}^{1,-1} + \partial^{0,-2} G^{1,1} = 0, \mbox{(b)}
{\cal D}^{0,2}d^{-1,-1}
+ i\partial_t{g}^{-1,1} + \partial^{-2,0} G^{1,1} = 0 \label{E}
\eea
where
\be
F \equiv \varepsilon^{\und{i}\und{k}}F_{\und{i}\und{k}}\,, \quad G^{1,1} \equiv
f^{(1,1)}\left(2 g^{1,-1}g^{-1,1} -\frac{1}{4}F^{\und{i}\und{k}}{}F_{\und{i}\und{k}}\right)
\ee
and
\bea
&& \nabla^{2,0} = \partial^{2,0} + (f^{1,1})^2 \partial^{0,-2}\,, \quad
\nabla^{0,2} = \partial^{0,2} + (f^{1,1})^2 \partial^{-2,0}\,, \nn \\
&& {\cal D}^{2,0} = \nabla^{2,0} + 2 (f^{1,1}\partial^{0,-2}f^{1,1})\;, \quad
{\cal D}^{0,2} = \nabla^{0,2} + 2 (f^{1,1}\partial^{-2,0}f^{1,1})\;.
\eea

First we solve eqs. \p{A1}. One can check that they imply the integrability condition
\be
\partial^{2,0}(f^{1,1}\partial^{-2,0}f^{1,1}) =
\partial^{0,2}(f^{1,1}\partial^{0,-2}f^{1,1})\,,
\ee
whence it follows that
\be
f^{1,1}\partial^{0,-2}f^{1,1} = \partial^{2,0}\varphi\,, \quad
f^{1,1}\partial^{-2, 0}f^{1,1} = \partial^{0,2}\varphi\,, \label{InT}
\ee
where $\varphi$, for the time being, is an arbitrary function of 
$t, u^{\pm 1}_i$ and $v^{\pm 1}_a\,$. Defining
\be
f^{1,1} = e^{-\varphi}\hat{f}^{1,1}\,,
\ee
one observes that \p{A} and \p{InT} together imply the linear equations for
$\hat{f}{}^{1,1}$
\be
\partial^{2,0}\hat{f}^{1,1} =   \partial^{0,2}\hat{f}^{1,1} = 0\, \quad \Rightarrow \quad
\hat{f}^{1,1} = f^{ia}(t)u^1_iv^1_a\,.
\ee
Then \p{InT} become equations for expressing $\varphi$ in terms of $f^{ia}$
\bea
&& \partial^{2,0}\varphi = e^{-2\varphi}(\hat{f}^{1,1}\hat{f}^{1,-1}) -
(\hat{f}^{1,1}\hat{f}^{1,1})e^{-2\varphi}\partial^{0,-2}\varphi\,, \nn \\
&& \partial^{0,2}\varphi = e^{-2\varphi}(\hat{f}^{1,1}\hat{f}^{-1,1}) -
(\hat{f}^{1,1}\hat{f}^{1,1})e^{-2\varphi}\partial^{-2,0}\varphi\,.
\eea
{}From these equations and the property that $\varphi$ is neutral
it follows that the harmonic variables can appear in $\varphi$ only through
the products $\hat{f}^{1,1}\hat{f}^{-1,-1}$ or  $\hat{f}^{1,-1}\hat{f}^{-1,1}\,$.
Because of the relation
\be
\hat{f}^{1,1}\hat{f}^{-1,-1} - \hat{f}^{1,-1}\hat{f}^{-1,1} = \frac{1}{2}f^{ia}f_{ia}
\equiv \frac{1}{2} f^2
\ee
only one of these products is independent. Choosing $X= \hat{f}^{1,1}\hat{f}^{-1,-1}$
as the argument of $\varphi$ we obtain for $\varphi$ the ordinary first-order
differential equation the general (regular in the limit $f^{ia} = 0$) solution of which
is
\be
f^{1,1} = \frac{2}{1 + \sqrt{1 + 4 X}}\hat{f}{}^{1,1}\,, \quad
X = \hat{f}^{1,1}\hat{f}^{-1,-1}\,, \label{Solvf}
\ee
where we absorbed a harmonic-independent  integration constant into rescaling of $f^{ia}$.
It is easy to directly check that \p{Solvf} solves both eqs. \p{A1}. 
From \p{Solvf} useful relations follow
\bea
&& \partial_t{f}{}^{1,1} =  \frac{1}{\sqrt{1 + 4 X}}\partial_t{\hat{f}}{}^{1,1} -
\frac{4}{(1 + \sqrt{1 + 4 X})^2\sqrt{1 + 4 X}}(\hat{f}^{1,1})^2\partial_t{\hat{f}}{}^{-1,-1}\,,
\nn \\
&& \partial^{0,-2} f^{1,1} =   \frac{1}{\sqrt{1 + 4 X}}\hat{f}{}^{1,-1}\,, \quad
\partial^{-2,0} f^{1,1} =   \frac{1}{\sqrt{1 + 4 X}}\hat{f}{}^{-1,1}\,. \label{Conseq}
\eea

The general scheme of solving the remaining harmonic equations \p{B} - \p{E} is as follows.
Taking into account that the fields $g^{1,-1}, g^{-1,1}$ and $F_{\und{i}\und{k}}$
have the dimension 1 and $d^{-1,-1}$ dimension 2, one expands them over all possible
structures having these dimensions, with the coefficients being functions of the
harmonic argument $X$. Then eqs. \p{B} - \p{E} amount to some first-order
inhomogeneous differential equations for these coefficients. The field
$F_{\und{i}\und{k}}$ is neutral, so eqs. \p{D} fix it up to the integration
constants $A_{(\und{i}\und{k})}(t), A(t)$ which are just the auxiliary fields of the
nonlinear ${\bf (4,8,4)}$ multiplet under consideration. The remaining fields
carry at least one negative harmonic $U(1)$ charge, so the corresponding equations
fully define them in terms of the fields $f^{ia}(t)$ and $A_{(\und{i}\und{k})}(t),
A(t)\,$.

In this way we obtain
\bea
&& F_{(\und{i}\und{k})} =   \frac{1}{\sqrt{1 + 4 X}} A_{(\und{i}\und{k})}\,, \quad
F = 2 A -8i\frac{1}{(1 + \sqrt{1 + 4 X})\sqrt{1 + 4 X}}\,
\partial_t{\hat{f}}{}^{-1,-1}\hat{f}^{1,1}\,, \label{A} \\
&& g^{-1,1} =  \frac{1}{\sqrt{1 + 4 X}}\left(A \hat{f}^{-1,1}
- i\partial_t{\hat{f}}{}^{-1,1} \right), \quad  g^{1,-1} =
\frac{1}{\sqrt{1 + 4 X}}\left(A \hat{f}^{1,-1}
- i\partial_t{\hat{f}}{}^{1,-1} \right). \label{Ag}
\eea
The expression for $d^{-1,-1}$ is rather cumbersome since there are quite a few possible
structures of dimension 2. It is convenient to represent it as a sum of six terms
\be
d^{-1,-1} = \sum_{\alpha =1}^6 d^{-1,-1}_\alpha\,,
\ee
and to give the solutions separately for each term
\bea
&& d^{-1,-1}_1 = \frac{1+ 2 f^2}{2(1 + 4 X)^{3/2}} A^2 \hat{f}^{-1,-1}\,, \quad
 d^{-1,-1}_2 =  \frac{1}{4(1 + 4 X)^{3/2}}\,(A^{(\und{i}\und{k})}A_{(\und{i}\und{k})})
\hat{f}{}^{-1,-1}\,, \nn \\
&&  d^{-1,-1}_3 = -i A \left[\frac{1}{\sqrt{1 + 4 X}}\partial_t{\hat{f}}{}^{-1,-1} -
\frac{2}{(1 + 4 X)^{3/2}}\left(\partial_t{\hat{f}}{}^{1,-1}\hat{f}{}^{-1,1}
+ \partial_t{\hat{f}}{}^{-1,1}\hat{f}{}^{1,-1} \right)\hat{f}{}^{-1,-1}\right], \nn \\
&& d^{-1,-1}_4 = \frac{1}{(1 + 4 X)^{3/2}}\left[ (\partial_t{\hat{f}}{}^{-1,-1})^2\hat{f}{}^{1,1}
+  \partial_t{\hat{f}}{}^{1,-1} \partial_t{\hat{f}}{}^{-1,1}\hat{f}{}^{-1,-1}\right], \nn \\
&& d^{-1,-1}_5 = -\frac{1}{\sqrt{1 + 4 X}} \partial_t^2{\hat{f}}{}^{-1,-1}\,, \quad
d^{-1,-1}_6 = -\frac{i}{\sqrt{1 + 4 X}} \partial_t{A} \hat{f}{}^{-1,-1}\,. \label{d}
\eea

\subsection{The bosonic sector of the bilinear action}
We will start with a sum of the simplest sigma-model type action \p{NONL2}
and the potential
term \p{Pot}. After doing the integration over $\theta$ s, the bosonic part of this action
is reduced to
\be
S^{bos}_2 = \int dt du dv \left(2 f^{1,1}d^{-1,-1} + 2 g^{-1,1}g^{1,-1}
- \frac{1}{4}\,F^{(\und{i}\und{k})} F_{(\und{i}\und{k})}
+\frac{1}{8} F^2 + \frac{i}{2}F\right) \equiv \int dt du dv {\cal L}_2.  \label{2comp}
\ee
Now one should substitute the expressions \p{Solvf}, \p{A}, \p{Ag} and \p{d} for the fields
in \p{2comp}. The resulting Lagrangian is rather simple
\be
{\cal L}_2 = \frac{1}{(1 + 4X)^{3/2}}\left(\partial_t{f}{}^{ia}\partial_t{f}_{ia} -
\frac{1}{4} A^{(\und{i}\und{k})} A_{(\und{i}\und{k})} -\frac{1}{2}(1 + 2 f^2) A^2 +
2i A\, f^{ia}\partial_t{f}_{ia} \right) + im A\,. \label{2fin1}
\ee
In writing the last term, we used the obvious property that the second
term in the expression for  $F(t,u,v)$ in \p{A} yields a total $t$-derivative
after integrating over harmonics. 

It remains to compute the bi-harmonic integral
\be
I = \int du dv \frac{1}{(1 + 4X)^{3/2}}\,.
\ee
It can be reduced, by rescaling $X \rightarrow \alpha X$,
and differentiating with respect to the parameter $\alpha$, to the simpler integral
\bea
I' =  \int du dv \sqrt{1 + 4X}\,, \nn
\eea
which can be computed, e.g.,  by expanding the integrand in powers of $X =
f^{ia}f^{kb}u^1_iu^{-1}_k v^1_av^{-1}_b\,$. The answer is
\be
I' = \frac{2}{3}\left(1 + \frac{1 + 2 f^2}{1 + \sqrt{1 + 2f^2}}\right)\,, 
\quad I = \frac{2}{(1 + \sqrt{1 + 2f^2})\sqrt{1 + 2f^2}}\,. \label{Iexpr}
\ee

Using this result and eliminating the auxiliary fields in \p{2fin1},
we arrive at the following
sigma-model type Lagrangian for the physical bosonic fields $f^{ia}(t)$:
\bea
{\cal L}_2 &=& \frac{2}{(1 + \sqrt{1 + 2f^2})\sqrt{1 + 2f^2}}\left[
\partial_t{f}{}^{ia}\partial_t{f}_{ia} - \frac{2}{1 + 2 f^2}\,(f^{ia}\partial_t{f}_{ia})^2\right] \nn \\
&& -\,\frac{m^2}{4}\,\frac{1 + \sqrt{1 + 2 f^2}}{\sqrt{1 + 2f^2}} -
\frac{2m}{1 + 2 f^2}\,(f^{ia}\partial_t{f}_{ia})\,.\label{2phys}
\eea
The last term is a total $t$-derivative and can be omitted. Note that the second part
of the target space metric within square brackets and non-trivial
potential term in \p{2phys}
resulted from the elimination of the auxiliary field $A(t)\,$.
Thus, starting from the bilinear superfield action \p{NONL2} and the linear superfield
potential term \p{Pot} we finally arrived at the non-trivial sigma-model Lagrangian
\p{2phys} with some nonlinear scalar potential $\sim m^2\,$.

Although the metric in \p{2phys} looks not conformally flat, it takes the conformally flat form 
in the properly chosen coordinates (like any four-dimensional $SO(4)$ invariant metric, e.g. that 
on the 4-sphere $SO(5)/SO(4)$). These coordinates are defined by
\be
y^{ia} = \frac{2}{1 +\sqrt{1 + 2f^2}}\, f^{ia}\,.\label{y}
\ee
In these coordinates, the Lagrangian \p{2phys} takes the simpler form
\be
{\cal L}_2 = \frac{2}{2 + y^2}\,\partial_t{y}{}^{ia}\partial_t{y}_{ia} - m^2 \frac{1}{2 + y^2}\,.
\label{2physy}
\ee

It is instructive to compare the pullback metric in \p{2phys} and \p{2physy} with the
analogous metric on the 4-sphere $SO(5)/SO(4)$. The latter, in the $f$ and $y$ coordinates,
takes the form
\be
\frac{1}{1 + 2f^2}\left[(\partial_t{f})^2 - \frac{2}{1 + 2 f^2}\,(f \cdot \partial_t{f})^2\right] =
\frac{4}{(2 + y^2)^2} (\partial_t{y})^2\,, \label{FS}
\ee
where we used the obvious condensed notation. The sigma model Lagrangian \p{FS}
enjoys an additional invariance under the nonlinear $SO(5)/SO(4)$ transformations
\be
\delta f^{ia} = b^{ia} + 2 (b\cdot f) f^{ia} \quad \mbox{or} \quad
\delta y^{ia} = (1 - \frac{1}{2}y^2) b^{ia} + (b\cdot y) y^{ia}
\ee
(they can be readily derived from the superfield $SO(5)/SO(4)$ transformation
\p{CosetTran}, \p{HarmCoset}). Lacking of the full $SO(5)$ invariance in
\p{2phys}, \p{2physy} is just a manifestation of the fact that the superfield
actions \p{NONL2}, \p{Pot} do not respect the full superconformal symmetry
$OSp(4^\star|4)$, but possess only ${\cal N}{=}8, d{=}1$ Poincar\'e supersymmetry and
$SO(4)$ R-symmetry. As we shall prove, the $S^4$ Lagrangian \p{FS}
cannot be reproduced using the nonlinear ${\bf (4,8,4)}$ multiplet.

\subsection{The general action}
After substituting the $\theta$ expansion \p{Theta} and doing the Berezin integral,
the bosonic part of the component Lagrangian in the general sigma-model type
superfield action \p{NONL} takes the form
\bea
L^{gen} = \frac{\partial {\cal L}^{2,2}}{\partial f^{1,1}}\,d^{-1,-1} +
\frac{\partial^2 {\cal L}^{2,2}}{\partial f^{1,1}\partial f^{1,1}}\, g^{-1,1}g^{1,-1}
-\,\frac{1}{8}\frac{\partial^2 {\cal L}^{2,2}}{\partial f^{1,1}\partial f^{1,1}}
\left(F^{(\und{i}\und{k})} F_{(\und{i}\und{k})}- \frac{1}{2} F^2\right), \label{genLag}
\eea
where ${\cal L}^{2,2} = {\cal L}^{2,2}(f^{1,1}, u, v)\,$. The action \p{2comp}
is reproduced under the choice ${\cal L}^{2,2} = f^{1,1}f^{1,1}$. The potential
invariant \p{Pot} makes
the same contributon $2im F$, and we will add it in the end.

As the next step, we substitute the expressions \p{Solvf}, \p{A}, \p{Ag} and \p{d}
for the harmonic fields in \p{genLag} and, after doing some algebra and eliminating
the auxiliary fields $A$ and $A_{(\und{i}\und{k})}$ by their equations of motion,
find the following generalization of the Lagrangian \p{2phys} (where we took into account
also the contribution of the potential term):
\be
L^{gen} = {\cal F}\left[(\partial_t{f})^2 - \frac{2}{1 + 2 f^2}\,(f \cdot \partial_t{f})^2\right]
-\frac{m^2}{1 + 2 f^2}\,({\cal F})^{-1}\,. \label{genPhys}
\ee
Here
\bea
{\cal F} (f^{ia}) =
 \frac{1}{2}\int du dv \frac{1}{1 + 4 X}\left[\frac{\partial^2 {\cal L}^{2,2}}
{\partial f^{1,1}\partial f^{1,1}} - 2 \frac{\hat{f}{}^{-1,-1}}{\sqrt{1 + 4X}}
\frac{\partial {\cal L}^{2,2}}{\partial f^{1,1}}\right]
=  \frac{1}{2}\int du dv \,\frac{\partial^2 {\cal L}^{2,2}}
{\partial \hat{f}{}^{1,1}\partial \hat{f}{}^{1,1}}\,, \label{Fcal}
\eea
and the last equality follows from the relation \p{Solvf}. Thus we observe the remarkable
fact that both the target space metric and the scalar potential are defined by the
same function ${\cal F}(f^{ia})$ \p{Fcal} which is a nonlinear generalization of
the metric function \p{Scomp1} of the linear ${\bf (4,8,4)}$ multiplet sigma-model action.
Using the representation \p{Fcal}, after some work one can derive the generalization
of the Laplace equation \p{Lap} to the nonlinear case. It is as follows
\be
\left[\Delta + 10 (f\cdot\partial_f) + 2 (f\cdot \partial_f) (f\cdot \partial_f)
+ 12\right] {\cal F} = 0\,, \label{NonlLap}
\ee
where $\Delta = \frac{\partial^2}{\partial f^{ia} \partial f_{ia}}$ and
$ (f\cdot \partial_f) =
f^{ia}\frac{\partial}{\partial f^{ia}}$. The derivation of \p{NonlLap} is essentially
based on the fact that, as follows from \p{Solvf},  ${\cal L}^{2,2}$ does not involve 
$\hat{f}{}^{-1,1}$ and $\hat{f}{}^{1,-1}$ and obeys the condition of the ``covariant''
independence of $\hat{f}^{-1,-1}$:
\be
\frac{\partial {\cal L}^{2,2}}{\partial \hat{f}{}^{-1,-1}} +
\frac{4}{(1 + \sqrt{1 + 4X})^2}\,(\hat{f}{}^{1,1})^2
\frac{\partial {\cal L}^{2,2}}{\partial \hat{f}{}^{1,1}} = 0\,,
\ee
which is also a corollary of the relation \p{Solvf}. Rescaling $f^{ia}$ and choosing the
scale as a contraction parameter, it is easy to see that \p{NonlLap} goes over into
the four-dimensional Laplace equation \p{Lap} when this parameter goes to zero. It is
straightforward to check that the factor \p{Iexpr} provides a particular,
$SO(4)$ invariant,
solution of \p{NonlLap}. On the other hand, the factor $1/(1 + 2 f^2)$ appearing in the
$S^4$ sigma model Lagrangian  \p{FS} is not a solution of \p{NonlLap}.
This means that one cannot
find a superfield Lagrangian ${\cal L}^{2,2}$ which would give rise to the $SO(5)$
invariant action in the bosonic limit. Hence one cannot construct superconformally
invariant action on the basis of the nonlinear ${\bf (4,8,4)}$ multiplet alone.
Note that, by going to the coordinates \p{y}, the sigma model part of
the action \p{genPhys}
can be brought into the conformally flat form, despite the fact that in the case
of arbitrary function ${\cal L}^{2,2}$ the $SO(4)$ symmetry inherent to the particular
action \p{2phys} is definitely broken.

We point out that {\it any} choice of the analytic Lagrangian
${\cal L}^{2,2}(f, u, v)$ gives the metric function ${\cal F}(f^{ia})$ obeying
eq. \p{NonlLap}, like any choice of ${\cal L}^{2,2}$ \p{s1gen} in the case of
linear ${\bf (4,8,4)}$
multiplet yields  the metric function $G(f)$, eq. \p{Scomp1}, satisfying
the 4-dim Laplace equation \p{Lap}. Thus, in both
cases the analytic bi-harmonic Lagrangians ${\cal L}^{2,2}$ serve
as unconstrained potentials of
the relevant geometries. While the target geometry associated with the linear ${\cal N}{=}8$
multiplet ${\bf (4,8,4)}$ is the same as that associated with its $d=2$ cousin, twisted
${\cal N}{=} (4,4)$ multiplet, i.e. the HKT geometry, it is not clear which generalization of
this geometry we are facing in the considered case of nonlinear ${\bf (4,8,4)}$ multiplet.
The natural conjecture would be that this is a sort of quaternion-K\"ahler geometry with
torsion. The crucial difference of the nonlinear multiplet from the linear one is that
the former seems not to be obtainable by dimensional reduction from higher dimensions,
at least we do not know how to accomplish this.

\section{Concluding remarks}
In this paper we constructed the new multiplet of ${\cal N}{=}8, d{=}1$ supersymmetry, 
a nonlinear version of the multiplet ${\bf (4,8,4)}$ \cite{BIKL2}, using the 
manifestly  ${\cal N}{=}8$ supersymmetric off-shell approach of the bi-harmonic 
${\cal N}{=}8$ superspace \cite{BISu}. This new multiplet is described, like its linear 
counterpart, by the analytic harmonic superfield $q^{1,1}(\zeta, u, v)$, 
this time subjected to nonlinear harmonic constraints. We found the realization 
of the $d{=}1$ superconformal group $OSp(4^\star\vert 4)$ in the analytic 
bi-harmonic superspace and showed that it non-trivially mixes the coordinates 
with the superfield $q^{1,1}\,$. The latter is a Goldstone superfield and it can be 
regarded as an extra analytic coordinate extending the analytic superspace to 
an analytic coset supermanifold of the full supergroup $OSp(4^\star\vert 4)\,$. 
Though it is impossible to construct a superconformally invariant action of 
$q^{1,1}\,$, the actions respecting invariance under ${\cal N}{=}8, d{=}1$ 
supersymmetry still exist, and we constructed the most general action of this kind.
The bosonic target metric in it satisfies a nonlinear generalization \p{NonlLap} of 
the four-dimensional Laplace equation. The bi-harmonic approach 
automatically provides a general solution of this equation, with the analytic 
superfield Lagrangian density as a prepotential (it also fully specifies the scalar potential). 
It still remains to identify the corresponding geometry. The generic 
metric is conformally flat in proper coordinates. 
In the simplest $SO(4)$ invariant case it 
is some deformation of the metric on $S^4 \sim SO(5)/SO(4)\,$. It would be 
interesting to seek some other solutions of eq. \p{NonlLap}.

As one of the problems for the future study, it is interesting to construct combined 
actions including both linear and nonlinear $q^{1,1}$ multiplets and to study the corresponding 
invariant actions and their geometric properties. Another problem is to try to add 
to the nonlinear $q^{1,1}$ multiplet some extra ${\cal N}{=}8, d{=}1$ multiplet containing 
a dilaton among its components (e.g. the multiplets ${\bf (5,8,3)}$ or 
${\bf (3,8,5)}$ \cite{BIKLb,BIKL2}) and to construct a superconformally invariant action 
of the nonlinear multiplet  ${\bf (4,8,4)}\,$ in such a way. Some other multiplets of 
${\cal N}{=}8, d{=}1$ supersymmetry, e.g. ${\bf (8,8,0)}$ and 
${\bf (7,8,1)}$ \cite{BIKL2}, also admit a description as constrained bi-harmonic 
analytic superfields \cite{ILeSu}. So one can hope to find out nonlinear analogs of 
these multiplets, following the lines of this paper.                                         
\vspace{0.3cm}

\noindent{\bf Note added}. When this paper was almost ready for the submission to e-archive, 
there appeared a work of three authors \cite{tri} where basically the same results 
were obtained from a nonlinear realization of $OSp(4^\star\vert 4)\,$ 
on the standard constrained ${\cal N}{=}8$ and ${\cal N}{=}4$, $d{=}1$ superfields. 
In particular, eq. \p{NonlLap} was derived. It was also shown that in the variables 
\p{y} it is reduced to the 4-dim Laplace equation for some new scalar metric function.    

\section*{Acknowledgments}
I thank Stefano Bellucci and Alessio Marrani for a few discussions at the initial stage 
of this study 
in May of 2005 in Frascati. I also acknowledge 
a support from RFBR grant, project  No 06-02-16684, and a grant of the 
Heisenberg-Landau program. 
This work was finalized during my visit to Departament de Fisica Teorica of 
Universitat de Valencia. I thank Jos\'e-Adolfo de Azc\'arraga for the kind hospitality 
and Dima Sorokin for a useful conversation.

\end{document}